\documentclass[aps,prl,twocolumn,showpacs,preprintnumbers,amsmath,amssymb,superscriptaddress]{revtex4}%

\usepackage{graphicx}%
\usepackage{dcolumn}
\usepackage{amsmath}
\usepackage{color}
\usepackage{bm}

\begin{document}

\title{Two-gap superconductivity in Ba$_{1-x}$K$_x$Fe$_2$As$_2$: A complementary study of the  magnetic penetration depth by $\mu$SR and ARPES}

\author{R.~Khasanov}
 \email{rustem.khasanov@psi.ch}
 \affiliation{Laboratory for Muon Spin Spectroscopy, Paul Scherrer
Institut, CH-5232 Villigen PSI, Switzerland}
\author{D.\,V.~Evtushinsky}
\affiliation{Institute for Solid State Research, IFW Dresden, P.\,O.\,Box 270116, D-01171 Dresden, Germany}
\author{A.~Amato}
 \affiliation{Laboratory for Muon Spin Spectroscopy, Paul Scherrer
Institut, CH-5232 Villigen PSI, Switzerland}
\author{H.-H.~Klauss}
 \affiliation{IFP, TU Dresden, D-01069 Dresden, Germany}
\author{H.~Luetkens}
 \affiliation{Laboratory for Muon Spin Spectroscopy, Paul Scherrer
Institut, CH-5232 Villigen PSI, Switzerland}
\author{Ch.~Niedermayer}
 \affiliation{Laboratory for Neutron Scattering, Paul Scherrer Institute and ETH
 Z\"urich,CH-5232 Villigen PSI, Switzerland}
\author{B.~B\"uchner}
\affiliation{Institute for Solid State Research, IFW Dresden, P.\,O.\,Box 270116, D-01171 Dresden, Germany}
\author{G.~L.~Sun }
 \affiliation{Max-Planck-Institut f\"ur Festk\"orperforschung, Heisenbergstra{\ss}e 1, 70569 Stuttgart, Germany}
\author{C.~T.~Lin}
 \affiliation{Max-Planck-Institut f\"ur Festk\"orperforschung, Heisenbergstra{\ss}e 1, 70569 Stuttgart, Germany}
\author{J.~T.~Park}
 \affiliation{Max-Planck-Institut f\"ur Festk\"orperforschung, Heisenbergstra{\ss}e 1, 70569 Stuttgart, Germany}
\author{D.~S.~Inosov}
 \affiliation{Max-Planck-Institut f\"ur Festk\"orperforschung, Heisenbergstra{\ss}e 1, 70569 Stuttgart, Germany}
\author{V.~Hinkov}
 \affiliation{Max-Planck-Institut f\"ur Festk\"orperforschung, Heisenbergstra{\ss}e 1, 70569 Stuttgart, Germany}

\begin{abstract}
We investigate the magnetic penetration depth $\lambda$ in superconducting Ba$_{1-x}$K$_x$Fe$_2$As$_2$ ($T_{\rm c}\simeq32$~K) with muon-spin rotation ($\mu$SR) and angle-resolved photoemission (ARPES). Using $\mu$SR, we find the penetration-depth anisotropy $\gamma_\lambda=\lambda_c/\lambda_{ab}$ and the second-critical-field anisotropy $\gamma_{H_{\rm c2}}$ to show an opposite $T$-evolution below $T_c$. This dichotomy resembles the situation in the two-gap superconductor MgB$_2$. A two-gap scenario is also suggested by an inflection point in the in-plane penetration depth $\lambda_{\rm{ab}}$ around 7~K. The complementarity of $\mu$SR and ARPES allows us to pinpoint the values of the two gaps and to arrive to a remarkable agreement between the two techniques concerning the full $T$-evolution of $\lambda_{\rm{ab}}$. This provides further support for the described scenario and establishes ARPES as a tool to assess macroscopic properties of the superconducting condensate.
\end{abstract}

\maketitle


Much effort is devoted to the investigation of the manifestations and the mechanism of unconventional superconductivity in the iron-arsenides, since many of their features clearly set them apart from other superconductors. Ab-initio calculations, for instance, indicate that superconductivity originates in the $d$-orbitals of the Fe ion, which normally would be expected to be pair-breaking \cite{Cao08,Singh08}. Several disconnected Fermi-surface sheets contribute to the superconductivity, as revealed by angle-resolved photoemission spectroscopy (ARPES) \cite{Ding08,Zhao08,Zabolotnyy08,Evtushinsky08}. Furthermore, indication for multi-gap superconductivity was obtained in measurements of the first and second critical fields $H_{\rm c1}$ and $H_{\rm c2}$ \cite{Ren08_2,Hunte08}, the magnetic penetration depth $\lambda$ \cite{Malone08,Hiraishi08}, as well as in point-contact Andreev reflection spectroscopy experiments \cite{Szabo08}.

Measurements of  $\lambda$ provide a conclusive method to reveal multi-gap superconductivity, since the presence of gaps with different gap-to-$T_{\rm c}$ ratios induces the appearance of inflection points in $\lambda(T)$ \cite{Carrington03,Khasanov07_La214,Khasanov07_Y124,Khasanov07_Y123}. Measuring $\lambda_i$ ($i=a$, $b$, or $c$) along certain crystallographic directions allows, in addition, the investigation of the penetration-depth anisotropy $\gamma_\lambda$. Within the London approximation, which implies $\lambda_i^{-2}\propto n_s/m^\ast_i$ ($n_s$ is the carrier concentration), $\gamma_\lambda$ is directly related to the anisotropy of the supercarrier mass $m^\ast$ via $\gamma_\lambda=\lambda_j/\lambda_i=\sqrt{m^\ast_j/m^\ast_i}$. As shown for the case of MgB$_2$ \cite{Angst04}, a different temperature evolution of $\gamma_\lambda$ and the second critical field anisotropy $\gamma_{H_{\rm c2}}$ is also indicative of multi-gap superconductivity.

Here we report a combined study of the penetration depth in a single crystal of Ba$_{1-x}$K$_x$Fe$_2$As$_2$ (BKFA) by means of $\mu$SR and ARPES. The sample was extensively characterized and several publications report its investigation by ARPES, magnetic neutron scattering, $\mu$SR and magnetic force microscopy \cite{Zabolotnyy08,Evtushinsky08,Park08}. Resistivity and dc-susceptibility measurements demonstrate a sharp superconducting (SC) transition at $T_{\rm c}=(32\pm1)$~K, reproducible among different crystals from the same growth batch, and X-ray powder diffraction has established the phase purity \cite{Park08}. Most importantly for our study, the gap structure was investigated by ARPES \cite{Evtushinsky08}. Furthermore, the occurrence of electronic phase separation into antiferromagnetic (AF) and superconducting/normal state regions on a lateral scale of several tens of nanometers was established \cite{Evtushinsky08,Park08}.


We begin by reporting on the $\mu$SR measurements which were carried  out at the $\pi$M3 beam line at the Paul Scherrer Institute (Villigen, Switzerland). The Ba$_{1-x}$K$_x$Fe$_2$As$_2$ single crystal with an approximate size of 5$\times$10$\times$0.06~mm$^3$ was mounted on a holder specially designed to perform $\mu$SR measurements on thin single-crystalline samples. The transverse-field (TF) and the zero-field (ZF) $\mu$SR experiments were performed at temperatures ranging from 1.5 to 200~K. In two sets of TF measurements the magnetic field ($\mu_0H=10$~mT~$>H_{\rm c1}$) was applied in parallel and perpendicularly to the crystallographic $c$-axis, respectively, and always perpendicularly to the muon-spin polarization. The typical counting statistics were $\sim10^{7}$ positron events for each particular data point.

\begin{figure}[tb]
\includegraphics[width=0.65\linewidth, angle=270]{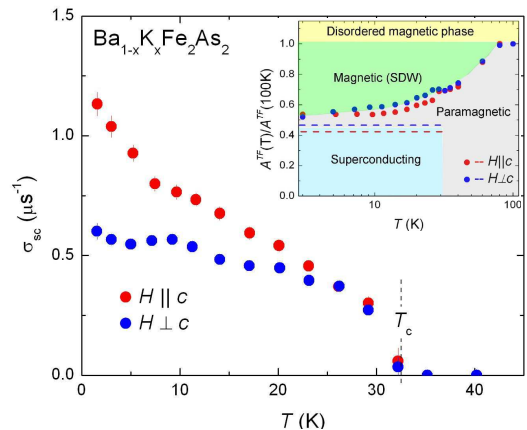}
%
\caption{(Color online) Temperature evolution of $\sigma_{\rm
sc}$  measured
after field cooling in $\mu_0H=10$~mT applied in parallel (red
circles) and perpendicularly (blue circles) to the crystallographic
$c$-axis. The inset shows the temperature evolution of the
initial TF asymmetry $A^{\rm TF}(t=0)=A_1+A_2$ (closed symbols)
and the superconducting asymmetry $A_1$ (dashed lines) normalized
to $A^{\rm TF}$ at $T=100$~K. The colored areas represent volume
fractions. Note the logarithmic $T$-scale.}
 \label{fig:sigma_ab-c}
\end{figure}

Experiments  in transverse field allow to study the magnetic ordering as well as to obtain the superfluid density response \cite{Goko08}. Muons stopping in magnetically ordered parts of the sample lose their  polarization relatively fast, since the magnetic field on the muon stopping site becomes a superposition of
the external and the internal fields. The superconducting response is observed as an additional damping below $T_{\rm c}$ because of the inhomogeneous field distribution of the external field penetrating the sample in form of vortices. Our ZF $\mu$SR experiments reveal that the signal from  the magnetically ordered parts vanishes within the first 0.3~$\mu$s. Bearing that in mind, in the whole temperature region the fit of TF data was restricted to times $t\geq0.3$~$\mu$s (see Ref.~\onlinecite{Goko08} for details). For $T<T_{\rm c}$, the TF $\mu$SR data were analyzed using the following two-component form:
\begin{eqnarray}
 A^{\rm TF}(t)&=&A_{1}\exp(-(\sigma_{\rm sc}^2+\sigma_{\rm nm}^2)t^2/2)\cos(\gamma
 B_{1}t+\phi)+  \nonumber \\
 &&A_{2}\exp(-\sigma_{\rm nm}^2t^2/2)\cos(\gamma
 B_{2}t+\phi).
 \label{eq:A_TF}
\end{eqnarray}
Here $A_1$ and $A_2$ are the initial asymmetries of the  first and the second component,
$\gamma/2\pi= 135.5$~MHz/T is the muon gyromagnetic ratio, $\phi$ is the initial phase of
the muon-spin ensemble, and the depolarization rates $\sigma_{\rm sc}$ and $\sigma_{\rm nm}$
characterize the damping due to the superconducting and the weak nuclear magnetic dipolar
contributions, respectively. The second term on the right-hand side of Eq.~(\ref{eq:A_TF})
accounts for the parts of the sample remaining in the normal state \cite{comment1}.
Each set of TF-$\mu$SR data, consisting of measurements in the $H\parallel c$ and $H\perp c$
configuration, respectively, was fitted simultaneously with $A_1$, $\sigma_{\rm nm}$, $\phi$,
and $B_2$ as common parameters and $\sigma_{\rm sc}$, $B_{1}$, and $A_2$ as individual parameters
for each temperature point.
The validity of our approach to fit some of the parameters globally was confirmed by
examining the evolution of $A_1$, $\sigma_{nm}$ and $B_2$ in the ``free'' fit.
%
Above $T_{\rm c}$, the fit was simplified to the single Gaussian component only, with
all parameters kept free. The results of the analysis are presented in Fig.~\ref{fig:sigma_ab-c}.

The inset in Fig.~\ref{fig:sigma_ab-c} shows that the initial TF asymmetry $A^{\rm TF}(t=0)=A_1+A_2$
(closed symbols) starts to decrease below $T\sim70$~K, following the gradual enhancement of the magnetic
fraction already investigated in this sample \cite{comment2,Park08}. Here, we concentrate on the SC
properties and study in detail the temperature range below $T_{\rm c}$, which remained unexplored in
the previous study \cite{Park08}. We note that $A^{\rm TF}$ (closed symbols) and the asymmetry related to the superconducting
fraction $A_1$ (dashed lines) are almost the same for the  $H\parallel c$ and $H\perp c$ sets of measurements.
This is exactly what is expected, since these asymmetries must represent the corresponding volume fractions
(magnetic or superconducting).

The temperature evolution of the SC part of the muon-spin depolarization rate $\sigma_{sc}$ is
presented in the main panel of Fig.~\ref{fig:sigma_ab-c}. It is worth to note that in a homogenous
superconductor $\sigma_{\rm sc}$ is expected to be proportional to the inverse squared magnetic penetration
depth, $\sigma_{\rm sc}\propto\lambda^{-2}$. In addition, in single-crystalline  sample the magnetic field
distribution in the superconductor in the mixed state is asymmetric and, therefore, can not be described by
a single Gaussian line (see e.g. \cite{Khasanov07_La214,Khasanov07_Y124,Khasanov07_Y123}). The
Ba$_{1-x}$K$_x$Fe$_2$As$_2$ sample studied here is, on the contrary, highly inhomogeneous \cite{Park08},
and the superconducting response, at our level of statistics, is well described by a single line of Gaussian shape.
We believe, however, that for Ba$_{1-x}$K$_x$Fe$_2$As$_2$ $\sigma_{\rm sc}$ is still a good measure of $\lambda$.
Indeed, $\sigma_{\rm sc}$ at $H\parallel c$ extrapolated to $T\rightarrow0$ results in
$\sigma_{\rm sc}(0)\simeq1.2$~$\mu$s$^{-1}$, which follows reasonably well the Uemura relation established
recently for various families of Fe-based superconductors \cite{Goko08,Luetkens08_Khasanov08_Aczel08}.
We conjecture that the antiferromagnetic islands act as preformed pinning centers for vortices, thus precluding
the formation of an ordered vortex lattice \cite{Park08}, while the screening current at this relatively low
field ($\mu_0H=10$~mT) still flows at a distance $\lambda$ from the vortex core.

Within the London model, the magnetic penetration depth  of the isotropic extreme type-II superconductor ($\lambda\gg\xi$, $\xi$ is the coherence length) is determined by $\lambda^{-2}\propto n_s/m^\ast$. For an anisotropic superconductor, the magnetic penetration depth is also anisotropic and is determined by an
effective mass tensor \cite{Thiemann89}:
\begin{equation} m^\ast_{\rm eff}=\left(
\begin{array}{ccc}
M_i & 0 & 0 \\
0 & M_j & 0 \\
0 & 0 & M_k \\
\end{array}
\right),
\end{equation}
where $M_i=m^\ast_i/\sqrt[3]{m^\ast_i \cdot m^\ast_j \cdot m^\ast_k}$ and $m^\ast_i$ is the mass of the carriers flowing along the $i$-th principal axis. The effective penetration depth for the magnetic field applied along the $i$-th principal axis of the effective mass tensor is then given by \cite{Thiemann89}:
\begin{equation}
\lambda_{\rm
eff}^{-2}=\frac{1}{\lambda_j\lambda_k}\propto\frac{1}{\sqrt{m_j^\ast
m_k^\ast}}\propto\sigma^{\parallel i}.
 \label{eq:lambda_jk}
\end{equation}
For convenience, we drop the index ``sc'' in the ``superconducting'' Gaussian relaxation rate $\sigma_{\rm sc}$. Equation~(\ref{eq:lambda_jk}) implies that by applying the magnetic field along the crystallographic $a$, $b$, and $c$ directions, one measures $\sigma^{\parallel a}\propto1/\lambda_b\lambda_{c}$, $\sigma^{\parallel b}\propto1/\lambda_a\lambda_c$ and $\sigma^{\parallel c}\propto1/\lambda_a\lambda_b$, respectively. By neglecting the difference between $\lambda_a$  and $\lambda_b$ the penetration depth anisotropy can be obtained as:
\begin{equation}
\gamma_\lambda=\lambda_c/\lambda_{ab}=\sigma^{\parallel{c}}/\sigma^{\perp{
c}}.
 \label{eq:gamma}
\end{equation}
%

\begin{figure}[tb]
\includegraphics[width=0.8\linewidth]{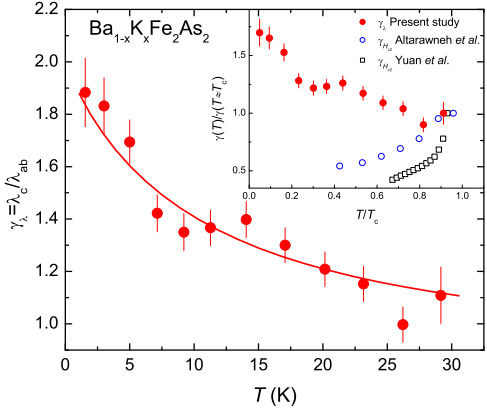}
%
\caption{(Color online) Temperature evolution of the magnetic
penetration depth anisotropy
$\gamma_\lambda=\lambda_c/\lambda_{ab}=\sigma^{\parallel{\rm
c}}/\sigma^{\perp{\rm c}}$. The line is a guide to the eye. In the
inset we compare $\gamma_\lambda$ with the $H_{\rm c2}$-anisotropy
$\gamma_{H_{\rm c2}}$ obtained for Ba$_{1-x}$K$_x$Fe$_2$As$_2$ by
Yuan {\it et al.} \cite{Yuan08} and Altarawneh {\it et al.}
\cite{Altarawneh08}, albeit in samples with somewhat different
$T_{\rm c}$. For better comparison, $T$ is divided by the respective
$T_{\rm c}$ and the values of $\gamma_\lambda$ and $\gamma_{H_{\rm
c2}}$ are normalized to 1 for the data point closest to $T_{\rm c}$.
}
 \label{fig:gamma}
\end{figure}

The temperature evolution of $\gamma_\lambda$ is presented  in Fig.~\ref{fig:gamma}. The inset shows the second critical field anisotropy $\gamma_{H_{\rm c2}}=H_{\rm c2}^{\perp{c}}/H_{\rm c2}^{\parallel{\rm c}}$ obtained for similar Ba$_{1-x}$K$_x$Fe$_2$As$_2$ samples in resistivity \cite{Yuan08} and
radio frequency penetration depth measurements \cite{Altarawneh08}. Within the phenomenological Ginzburg-Landau theory, in a single-gap superconductor both anisotropies  must be equal \cite{Kogan81}:
\begin{equation}
\gamma_\lambda=\frac{\lambda_c}{\lambda_{ab}}=
\sqrt{\frac{m^\ast_c}{m^\ast_{ab}}}= \gamma_{H_{\rm
c2}}=\frac{H_{\rm c2}^{\perp{c}}}{H_{\rm c2}^{\parallel{c}}}=
\frac{\xi_{ab}}{\xi_c}.
 \label{eq:anisotropy}
\end{equation}
It is natural to expect that the values of the same  quantities measured by various techniques should be the same. While a deviation at a particular temperature might be explained by a slight variation of the properties among samples used in the different experiments, this cannot account for the opposite temperature evolution of $\gamma_\lambda$ and $\gamma_{H_{\rm c2}}$ shown in the inset of Fig.~\ref{fig:gamma}. Hence, in Ba$_{1-x}$K$_x$Fe$_2$As$_2$ $\gamma_\lambda$  and $\gamma_{H_{\rm c2}}$ are {\it not the same} and Eq.~(\ref{eq:anisotropy}) is violated. This resembles the situation in the two-gap superconductor MgB$_2$, albeit with reversed trends for $\gamma_\lambda$ and $\gamma_{H_{\rm c2}}$: In MgB$_2$, $\gamma_\lambda$ was found to decrease with decreasing temperature from about 2 to 1.1, while $\gamma_{H_{\rm c2}}$ increases from $\sim$2 at $T_{\rm c}$ to 6 at low temperatures \cite{Angst04}. It is worth noting that recently the presence of two distinct anisotropies $\gamma_\lambda$ and $\gamma_{H_{\rm c2}}$ was reported
for the single-layer MeFeAsO$_{1-x}$F$_x$ (Me=Nd and Sm) \cite{Weyeneth08} and the double-layer BaFe$_{2-x}$Co$_x$As$_2$ \cite{Tanatar09}. The authors of Ref. \onlinecite{Weyeneth08} also explain the observed behavior by the presence of multiple gaps opening on various bands at the Fermi level.

\begin{figure}[b]
\includegraphics[width=1.0\linewidth, angle=0]{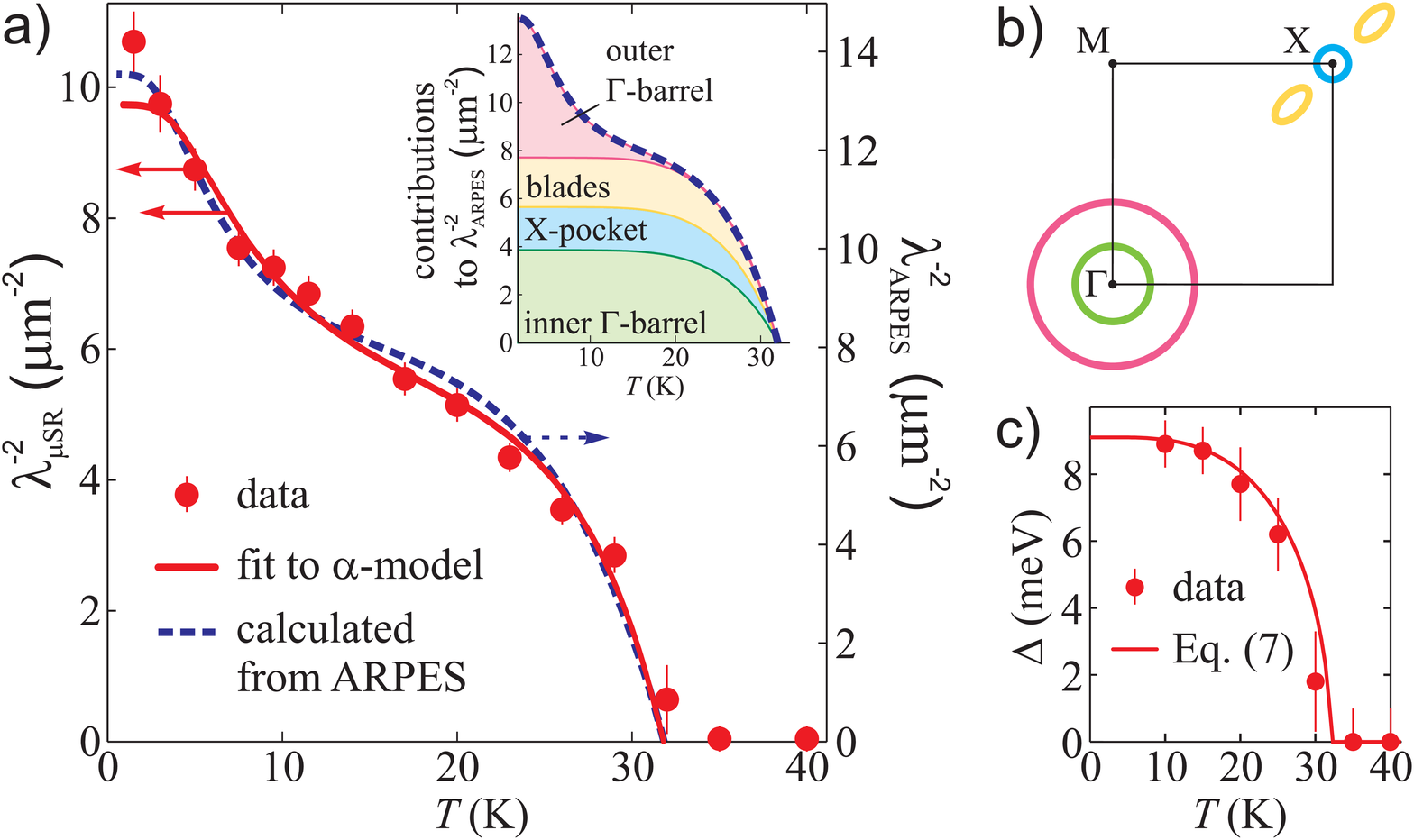}
 \vspace{-0.5cm}
\caption{(Color online) Temperature evolution of the inverse
squared in-plane magnetic penetration depth $\lambda_{ab}^{-2}$ obtained from the measured
$\sigma^{\parallel c}$ presented in Fig.~\ref{fig:sigma_ab-c} by
using the relation $\sigma(\mu{\rm
s}^{-1})=0.1067\lambda^{-2}(\mu{\rm m}^{-2})$ \cite{Brandt88}. The
solid line represents the result of a fit of
Eq.~(\ref{eq:lambda_ab}) to the $\alpha$-model, the dashed line represents a calculation of $\lambda_{ab}^{-2}$ from the electronic structure revealed by ARPES \cite{Evtushinsky2} with \emph{one} fitting parameter, $\Delta_{2}$. Inset: contributions of different Fermi surface sheets to $\lambda_{ab}^{-2}$. (b) Fermi surface of BKFA. (c) Temperature dependence of the SC gap, extracted from ARPES spectra \cite{Evtushinsky08}.}
 \label{fig:lambda_ab}
\end{figure}

An additional confirmation for the multi-gap  behavior comes from
the analysis of the temperature dependence of the in-plane magnetic
penetration depth. Fig.~\ref{fig:lambda_ab} shows
$\lambda_{ab}^{-2}(T)$ obtained from the measured
$\sigma^{\parallel{c}}(T)$ by using the relation $\sigma(\mu{\rm
s}^{-1})=0.1067\cdot\lambda^{-2}(\mu{\rm m}^{-2})$ \cite{Brandt88}.
The experimental $\lambda_{ab}^{-2}(T)$ data were analyzed within
the framework of the phenomenological $\alpha$-model by assuming two
independent contributions to the total $\lambda_{ab}^{-2}$ \cite{Carrington03}:
\begin{eqnarray}
\lambda_{ab}^{-2}(T)&=& \lambda_{ab}^{-2}(0)\left( \omega\cdot \{1-D[\Delta_1(T), T]\}\right. \nonumber \\
&&\left. +(1-\omega)\cdot\{1- D[\Delta_2(T), T]\} \right),
 \label{eq:lambda_ab}
\end{eqnarray}
where $D(\Delta, T) \equiv \int_{-\infty}^{+\infty} \left(-\frac{\partial f(\varepsilon)}{\partial \varepsilon}\right){\rm Re}\frac{\varepsilon}
{\sqrt{\varepsilon^2-\Delta^2}}{\rm d}\varepsilon$ \cite{Evtushinsky2}, $f(\varepsilon)$ is the Fermi function, and $\omega$ is the contribution of the bigger gap to $\lambda_{ab}^{-2}$.
The temperature dependence of the superconducting gap is assumed to be \cite{Carrington03}
\begin{equation}
\Delta_{1,2}(T)=\Delta_{1,2}(0)\tanh\{1.82[1.018(T_{\rm c}/T-1)]^{0.51}\},
    \label{eq:gapTdep}
\end{equation}
in agreement  with $\Delta(T)$ measured by ARPES (see Fig.~\ref{fig:lambda_ab}c).
The solid line in Fig.~\ref{fig:lambda_ab}a represents the result of a fit of Eq.~(\ref{eq:lambda_ab}) to the experimental data with $\lambda_{ab}(0)$, $\omega$, $\Delta_1$, and $\Delta_2$ as free parameters. The fit yields: $\Delta_1=9.1$~meV, $\Delta_2=1.5$~meV, $\omega=0.5$, and $\lambda_{ab}(0)=320$~nm \cite{comment3}.
%

The penetration depth $\lambda_{ab}(T)$ can also be calculated from the electronic band dispersion and the momentum-resolved SC gap \cite{Chandrasekhar} which were determined by ARPES on BKFA single crystals from the same growth batch \cite{Zabolotnyy08, Evtushinsky08, Volodya2, Evtushinsky2}. The Fermi surface consists of four different sheets\,---\, an inner $\Gamma$-barrel, an X-pocket and blade-shaped pockets with a large isotropic gap $\Delta_1$, and an outer $\Gamma$-barrel with a small gap $\Delta_2$ \cite{Zabolotnyy08, Volodya2} (Fig.~\ref{fig:lambda_ab}b). The formula relating $\lambda$ to the electronic structure reads \cite{Evtushinsky2}

\begin{equation}
\lambda_{ab}^{-2}(T) = I_1\left\{ 1 - D[\Delta_1(T), T] \right\} + I_2\left\{ 1 - D[\Delta_2(T), T]\right\},
 \label{eq:lambda_ARPES}
\end{equation}
where $I_{1,2}$ are integrals over the Fermi-surface contours
\begin{equation}
I_1 = \frac{e^2}{2\pi\varepsilon_0c^2hL_c}\!\!\!\!\!\!\!\! \int\limits_{\begin{smallmatrix} \text{outer }\Gamma, \\ \text{blades} \\ \text{X-pocket}
\end{smallmatrix}}\!\!\!\!\!\!\! v_\text{F}(\mathbf{k}) {\rm d}k, \ \  I_2 = \frac{e^2}{2\pi\varepsilon_0c^2hL_c}\!\!\!\!\!\! \int\limits_{\text{inner }\Gamma}\!\!\!\! v_\text{F}(\mathbf{k}) {\rm d}k,
\end{equation}
$\varepsilon_0$, $h$, $e$, $c$ are physical constants, $L_c$ is the $c$-axis lattice parameter, and $v_\text{F}$ is the Fermi velocity. Further details of the electronic structure and the calculation are given in ref. \onlinecite{Evtushinsky2}. Eq.~\eqref{eq:lambda_ARPES} is equivalent to Eq.~(\ref{eq:lambda_ab}) with $\lambda_{ab}^{-2}(0) = I_1 + I_2$ and $\omega = \frac{I_1}{I_1+I_2}$. Using Eq. \eqref{eq:lambda_ARPES}, we calculate $\lambda_{ab}^{-2}(0)$ and $\omega$, while $\Delta_1$ is known from the ARPES measurements \cite{Zabolotnyy08, Evtushinsky08}. A comparison of these parameters determined by the two different methods is shown in Table~I. Taking into account the complementarity of the methods, the agreement is remarkable and strengthens the validity of the obtained results. The discrepancy in $\lambda_{ab}(0)$ is well within the range of values obtained in different $\mu$SR experiments \cite{Hiraishi08}.

Now we can assess the remaining parameter $\Delta_2$ by fitting Eq.~\eqref{eq:lambda_ARPES} to the
measured $\lambda_{ab}(T)$ normalized to  $\lambda(0)$. In Fig.~\ref{fig:lambda_ab} the normalization is
realized by scaling the $\lambda^{-2}_{\scriptscriptstyle{\mu \rm{SR}}}$- and
$\lambda^{-2}_{\scriptscriptstyle{\rm{ARPES}}}$-axes accordingly. We obtain $\Delta_2=1.1$~meV, again
in good agreement with the $\mu$SR result.

\begin{table}[]
\begin{tabular}{l@{~~~}|l@{~~~~~}l@{~~}}
                   &$\mu$SR&ARPES\\
\hline
$\lambda_{ab}(0)$\,(nm) & 320   & 270 \\
$\omega$           & 0.51  & 0.55 \\
$\Delta_1$\,(meV)  & 9.1   & 9.1 \\
$\Delta_2$\,(meV)  & 1.5   & $<4$ \\
\end{tabular}
\caption{Parameters as extracted from the fit of $\mu$SR data and calculated from ARPES data.}
\label{tab1}
\end{table}

In summary, from $\mu$SR measurements on a single-crystalline sample of BKFA ($T_{\rm c}\simeq32$~K)
we have determined the anisotropy of the magnetic-field
penetration depth $\gamma_\lambda=\lambda_c/\lambda_{ab}$.
The penetration depth anisotropy increases with decreasing $T$ from $\gamma_\lambda\simeq1.1$ at
$T\simeq T_{\rm c}$ to $\gamma_\lambda\simeq1.9$ at $T\simeq1.7$~K, while the $T$-evolution of the
$H_{\rm c2}$ anisotropy $\gamma_{H_{\rm c2}}$ shows an opposite trend \cite{Yuan08,Altarawneh08}.
This resembles very much the situation in double-gap MgB$_2$ where both anisotropies are equal at
$T_{\rm c}$, but evolve oppositely with $T$. The notion of two SC gaps is supported by the
observation of an inflection point in $\lambda_{ab}$ at $\sim7$~K. From a fit of $\lambda^{-2}_{ab}$
to the phenomenological $\alpha$-model we obtain gap values of $\Delta_1=9.1$~meV and $\Delta_2=1.5$~meV.
A comparison of $\lambda_{ab}^{-2}(T)$ measured by $\mu$SR with the one calculated from ARPES data shows a
remarkable agreement between these two complementary approaches, lending further support to our conclusions
and establishing ARPES as a tool to estimate $\lambda_{ab}$.


The $\mu$SR work was performed at the Swiss Muon Source (S$\mu$S),  Paul Scherrer Institute (PSI, Switzerland). ARPES results were obtained at BESSY.


\begin{thebibliography}{99}
\bibitem{comment1} The second term in Eq.~(\ref{eq:A_TF}) may also include the background contribution
arising from the muons stopped outside the sample.
\bibitem{comment2} The difference between $A^{\rm TF}(T)$
obtained  in the present study to that reported in \cite{Park08}
is due to different normalization. Here $A^{\rm TF}(T)$ is
normalized to $A^{\rm TF}(T=100$~K), while in
Ref.~\onlinecite{Park08} $A^{\rm TF}(T)$ is normalized to the
maximum asymmetry ($\simeq21$\%) at our experimental setup.
\bibitem{comment3} We estimate the true overall uncertainty in the absolute gap values to be of the order of 0.5--1~meV.
\end{thebibliography}
\end{document}